\documentstyle[12pt]{article}
\def\nn{\nonumber}
\def\ee{\end{equation}}
\def\be{\begin{equation}}
\def\eea{\end{eqnarray}}
\def\bea{\begin{eqnarray}}
\def\eeas{\end{eqnarray*}}
\def\beas{\begin{eqnarray*}}
 
\begin{document}
\begin{titlepage}
\begin{center}
{\large\bf Bosonization of the pairing Hamiltonian}
\vspace*{1cm}

{M.B.Barbaro$^1$, A.Molinari$^1$, 
F.Palumbo$^2$ and M.R.Quaglia$^1$}

\vspace*{1cm}
{\it 
$^1$ Dipartimento di Fisica Teorica dell'Universit\`a di Torino
and INFN Sezione di Torino, via P.Giuria 1, I--10125, Torino, Italy\\
$^2$ INFN - Laboratori Nazionali di Frascati,P.O.Box 13, 
I--00044 Frascati, Italy}

\vspace*{2cm}

{\bf ABSTRACT}\\

\begin{quotation}
{\it The problem of the pairing interaction is dealt with even Grassmann 
variables in the hamiltonian framework . Eigenfunctions of given energy, 
seniority and zero third component of angular momentum are given in terms 
of single particle and collective bosonic fields.}

{\it Si tratta il problema dell'interazione di pairing con le 
va\-ria\-bi\-li di 
Grassman pari nel formalismo hamiltoniano. Si danno le autofunzioni di data 
energia, seniorit\`a e terza componente del momento angolare pari a zero,
in termini di campi bosonici di particella singola e collettivi.}
\end{quotation}
\end{center}
\end{titlepage}

Recently the problem of bosonizing the pairing hamiltonian $H_P$ has been 
addressed in the formalism of the path integrals over Grassmann variables 
\cite{Bar98}. 
In this context, using even elements of the Grassman algebra, 
a functional integral representation
of the correlation functions has been given by performing a 
change of variables in the Berezin integral.
It has thus been possible to express 
the pairing action and the ground state wave function
through ``collective variables'', namely specific 
linear combinations of the even elements 
of the algebra, and to obtain the ground state energy of $H_P$.

However, to solve the problem of the excited states of $H_P$ 
(with non zero seniority $v$) in the path integral framework 
has proved to be a quite difficult task to perform. 
Accordingly here we address the $v\ne 0$ problem resorting to 
the Grassmann algebra in the hamiltonian framework. 

Of course the spectrum of $H_P$ has already been found long time ago 
within the quasi spin formalism \cite{Row70}, 
but still we believe it useful to address the same problem in this new 
context to shed light on the
role of composite fields in nuclear physics and to provide 
explicit expressions for the eigenfunctions of 
$H_P$ with a non vanishing seniority.

To start with, 
we shortly recall the essential elements of the hamiltonian
many-fermions problem in terms of Grassmann variables (the bosonic 
problem, likewise, can be formulated in terms of 
holomorphic variables) \cite{Fad80}.

One exploits the isomorphism between the Fock space $F$ generated by the 
fermionic creation operators $\hat{a}_1^+,\cdots \hat{a}_{2\Omega}^+$ 
and the algebra ${\cal{G}}^+$ generated by the set of totally 
anticommuting objects $\lambda^*_1,...\lambda^*_{2\Omega}$ (odd 
elements of the algebra). The isomorphism is defined by mapping the vectors
$\hat{a}^+_1...\hat{a}^+_j|0>$ 
onto the elements 
$\lambda^*_1...\lambda^*_j$. 
The image of a generic vector $|\Psi> \in F$ under this mapping will be
denoted by $\Psi(\lambda^*)$.

The scalar product in $F$ in terms of Grassmann variables reads then

\be
<\Psi_1|\Psi_2> = \int d\lambda^*_{2\Omega} d\lambda_{2\Omega}...
d\lambda^*_1d\lambda_1 
(\Psi_1(\lambda^*))^*  \mu_-(\lambda^*\lambda) \Psi_2(\lambda^*), 
\ee
the measure being 

\begin{equation}
\mu_{\pm}(\lambda^*\lambda)=e^{\pm \sum_i \lambda^*_i \lambda_i}.
\end{equation}
Actually to define the kernels of integral operators the measure with the plus
sign is needed. 

Indeed a linear operator in normal form, namely
 
\begin{equation}
\hat{{\cal O}}=\sum_{i_1...i_k} \sum_{j_1...j_k} 
{\cal O}^{i_1...i_k,j_1...j_k}\hat{a}^+_{i_1}...\hat{a}^+_{i_k}
\hat{a}_{j_1}...\hat{a}_{j_k}~, 
\end{equation}
can be expressed in terms of Grassmann variables according to 

\begin{equation}
{\cal{O}}(\lambda^*,\lambda )=\sum_{i_1,...i_k,j_1...j_k}
{\cal{O}}^{i_1,...i_k,j_1...j_k}
\lambda^*_{i_1}...\lambda^*_{i_k}\lambda_{j_1}...\lambda_{j_k}~,
\end{equation}
and its associated kernel reads

\begin{equation}
K_{{\cal{O}}}(\lambda^*,\lambda)={\cal{O}}(\lambda^*,\lambda) 
\mu_+(\lambda^*\lambda).
\end{equation}

Then the action of ${\cal{O}}$ on a state $\Psi$ is given, in integral form, as

\begin{equation}
({\cal{O}} \Psi)(\lambda^*)= \int [d\lambda'^* d\lambda'] 
K_{\cal{O}}(\lambda^*,\lambda') \mu_-(\lambda'^*\lambda')
\Psi(\lambda'^*)\ .
\label{eq:Opsi}
\end{equation}

We apply the formalism, shortly revisited above, to
a system made of one kind of nucleons 
(e.g. neutrons) in a level specified by the angular momentum $j$, hence the 
indices labelling the $\lambda$ variables will correspond to 
the third components $m$ of $j$.

Now if only pairs of nucleons coupled to a total 
angular momentum with $J_z =0$ are considered, then
the exponentials appearing in $\mu_{\pm}$ can be transformed as follows

\be
\int [d\lambda^*] \exp\left(\pm \lambda^*_m \lambda_m \pm 
\lambda^*_{-m} \lambda_{-m}\right) f(\phi^*)=
 \int [d\phi^*]\exp\left(\phi^*_m \phi_m\right) f(\phi^*)
\label{eq:Opsi_2}
\ ,
 \ee
in terms of the {\it even Grassmann variables}

\be
\phi_m = \lambda_{-m} \lambda_m ~.
\ee

For $N$ (even) nucleons in a single particle level $j$, the pairing 
hamiltonian reads
\be
{\hat H}_P^{(j)} = -g_P {\hat A}^\dag_j {\hat A}_j \ ,
\label{H_pairing}
\ee
where
\be
{\hat A}_j= \sqrt{\frac{\Omega}{2}}
\sum_{m=-j}^j \langle j m, j -m |00\rangle {\hat a}_{j-m}{\hat a}_{jm} =
\sum_{m=1/2}^j (-1)^{j-m} {\hat a}_{j-m}{\hat a}_{jm} \ ,
\ee
being $\Omega =(2j+1)/2$ and $\langle j m, j -m |00\rangle$ the usual 
Clebsch Gordan coefficient. 
In the frame of the Grassmann algebra, we recast instead Eq.(\ref{H_pairing}) 
as follows 
\be
H_P^{(j)} = -g_P A^*_j A_j
\ee
with
\be
A_j= 
\sum_{m=1/2}^j (-1)^{j-m} \lambda_{-m} \lambda_m
=\sum_{m=1/2}^j (-1)^{j-m} \phi_m \ .
\ee

We explore now whether the $\phi$'s can be the building blocks
of the eigenstates of $H_P^{(j)}$. Indeed this turns out to be the case, since, 
according to (\ref{eq:Opsi}) and (\ref{eq:Opsi_2}), the action of 
the hamiltonian over states set up with the $\phi$'s is
\be
H_P^{(j)} \psi(\phi^*) = \int [d\eta^* d\eta] K_P^{(j)} 
(\phi^*,~ \eta) e^{\sum{\eta^*\eta}}
\psi(\eta^*) \ ,
\ee
the kernel reading
\be
K_P^{(j)}(\phi^*,~ \eta)= H_P^{(j)}(\phi^*,~ \eta) e^{\sum{\phi^*\eta}}~.
\ee
Hence the eigenvalues equation 
\be
H_P^{(j)} \psi(\phi^*) = \int [d\eta^* d\eta] K_P^{(j)} (\phi^*,~ \eta) 
e^{\sum{\eta^*\eta}}\psi(\eta^*) = E^{(j)} \psi(\phi^*)
\label{eq_eigenvalue}
\ee
follows.

We then look for eigenstates of $H_P^{(j)}$ 
in the form of linear combinations of products of $\phi$'s.
More specifically, for fixed $\Omega$ and for $n=N/2$ pairs,
the wave function will be a superposition of ${\Omega}\choose{n}$ monomials, 
of the type
\be
\psi (\phi^*) = \sum_{i=1}
^{{\Omega}\choose{n}} 
\beta_i {[\phi^*_{m_1}
\cdots \phi^*_{m_n}]}_i~,
\label{psi_generale}
\ee
where the $\beta_i$ are complex coefficients. 

In order to study the secular equation of $H_P^{(j)}$, we construct 
the matrix associated to the latter in the basis of the states 
(\ref{psi_generale}). 
To start with, we consider {\it one pair only}. In this case, the
wave function and the eigenvalue equation read, respectively, 
\be
\psi(\phi^*) = \sum_{m=1/2}^{j} (-1)^{j-m} \beta_m \phi^*_m
\label{basis}
\ee
and 
\be
({\cal E}+1) \beta_m +\sum_{p(\ne m)=1/2}^j \beta_p  =0
\label{eq_n1}
\ee
where ${\cal E}= E/g_P$ .
Moreover the representation of $H_P^{(j)}$ in the basis (\ref{basis}) is 
\be\left(
\begin{array}{cccccc}
{\cal E} +1 & 1 & 1 & \cdots & \cdots &1\\
1&{\cal E} +1 & 1  & \cdots & \cdots &1\\
1& 1&{\cal E} +1   & \cdots & \cdots &1\\
1&1&\cdots &{\cal E} +1& 1&1\\   
1&1&\cdots &\cdots &{\cal E} +1&1\\   
1&1& 1&\cdots & \cdots &{\cal E} +1   
\end{array} \right)~,
\label{matrix_n1}
\ee
a matrix of dimension $\Omega$ with 
the index $m$ labelling the rows and the columns. 
This matrix is obviously invariant for a permutation of the 
values of $m$. 
Elementary methods lead then to the characteristic equation 
$$\left( {\cal E}+ \Omega \right) {\cal E}^{\Omega-1} =0~.$$ 
We thus see that, of the $\Omega$ expected real roots, two only are 
distinct, namely the lower one 
$$ {\cal E}_0(n=1) =- \Omega$$
with multiplicity $\delta=1$ and 
$$ {\cal E}_2(n=1) =0$$
with multiplicity $\delta=\Omega-1$. 

In turn, the associated eigenvectors read, respectively,
\be
\Psi_0(n=1) = \frac{1}{\sqrt{\Omega}}\sum_{m=1/2}^j (-1)^{j-m} \phi^*_m
\label{psi0_n1}
\ee
and 
\be
\Psi_2(n=1)= {\cal N} \sum_{m=1/2}^{j-1}\beta_m \left[
(-1)^{j-m} \phi^*_m - \phi^*_j\right]~,
\label{psi2_n1}
\ee
${\cal N}$ being a normalization factor. 
The above findings agree with the physics of the hamiltonian $H_P^{(j)}$, 
which acts only in presence of $J=0$ pairs.
Also the eigenvectors (\ref{psi0_n1}, \ref{psi2_n1}) are orthogonal 
and the eigenvalues are those obtained in the usual quasi-spin framework. 

Of significance is that the characteristic equation
for the case with $n=\Omega-1$, namely
\be
({\cal E}+\Omega-1) \beta_m + \sum_{p(\ne m)=1/2}^j \beta_p  =0~,
\label{eq_nomega-1}
\ee
has the same structure of (\ref{eq_n1}). However the states read now
$$\psi(\phi^*)= 
\sum_{m=1/2}^j (-1)^{j-m} \beta_m \phi^*_{1/2} \cdots \phi^*_{m-1}\phi^*_{m+1}
\cdots \phi^*_{j} ~, $$
the index $m$ corresponding to the missing pair. 

By extension, one finds that the case with $n$ pairs present is equivalent 
to the one with $\Omega -n $ pairs: indeed it leads to the same matrix but 
for the diagonal elements. Hence in the following we shall confine ourselves 
to explore the cases with $n\le \frac{\Omega}{2}$ only.

Next we address the case $n=2$. Here the wave function reads
\be
\psi(\phi^*) = \sum_{m=1/2}^{j} \sum_{n=1/2}^{m-1}
(-1)^{j-m} (-1)^{j-n} \beta_{mn} 
\phi^*_m \phi^*_n \ ,
\ee
entailing the secular equation 
\be
({\cal E}+2) \beta_{mn}
 + \sum_{p(\ne n,m)=1/2}^j \left(\beta_{pm} + \beta_{pn} \right) =0 \ .
\label{eq_n2}
\ee
As for the case $n=1$, the associated matrix can be written in 
${{\Omega}\choose{2}}!$ equivalent ways (not all of them being
distinct), each one corresponding to a different labelling of the states,
but yielding the same determinant.

For example, for $\Omega=4$ (the smallest $\Omega$ for $n=2$)
one finds the following $6\times 6$ matrix
\be\left(
\begin{array}{cccccc}
{\cal E} +2 & 1 & 1 & 1 & 1 &0\\
1&{\cal E} +2 & 1  & 1 & 0 &1\\
1& 1&{\cal E} +2   & 0 & 1 &1\\
1&1&0&{\cal E} +2& 1&1\\   
1&0&1&1 &{\cal E} +2&1\\   
0&1& 1&1 & 1& {\cal E} +2   
\end{array}\right) \ ,
\label{matrix_n2omega4}
\ee
with the characteristic equation 
$$({\cal E}+6)({\cal E}+2)^3{\cal E}^2=0  ~.$$
Thus, in this case, out of 6 real roots only 3 turn out to be distinct, namely 
the lowest 
$$ {\cal E}_0(n=2)=-6$$
with degeneracy $\delta=1$, the intermediate
$$ {\cal E}_2(n=2)=-2$$ 
with degeneracy $\delta=3$ and the highest 
$$ {\cal E}_4(n=2)=0$$ 
with degeneracy $\delta=2$.
The associated orthogonal eigenvectors are found to be
\bea
&& \Psi_0(n=2)= 2 {\cal N}_0 
\left(-\phi^*_{1/2}\phi^*_{3/2}+ \phi^*_{1/2}\phi^*_{5/2}
- \phi^*_{1/2}\phi^*_{7/2} \right.
\nn\\
&&\left.\ \ \ \ \ \ \ \ -\phi^*_{3/2}\phi^*_{5/2}
+\phi^*_{3/2}\phi^*_{7/2}- \phi^*_{5/2}\phi^*_{7/2}\right)
\label{psi0_n2}\\
&& \Psi_2(n=2)= {\cal N}_2 \left[
a_1 \left(\phi^*_{1/2}\phi^*_{3/2}-\phi^*_{5/2}\phi^*_{7/2}\right) 
\right.\nn\\
&&\ \ \ \ \ \ \ \  + a_2 \left(\phi^*_{1/2}\phi^*_{5/2}-
\phi^*_{3/2}\phi^*_{7/2}\right) \nn\\
&& \ \ \ \ \ \ \ \  \left. + a_3 \left(\phi^*_{1/2}\phi^*_{7/2}-
\phi^*_{3/2}\phi^*_{5/2}\right) \right]\label{psi2_n2}\\
&& \Psi_4(n=2)={\cal N}_4 \left[
(b_2 - b_3)\left(\phi^*_{1/2}\phi^*_{3/2}+\phi^*_{5/2}\phi^*_{7/2}\right) 
\right. \nn\\
&& \ \ \ \  \ \ \ \ +  b_2 \left(\phi^*_{1/2}\phi^*_{5/2}+
\phi^*_{3/2}\phi^*_{7/2}\right) \nn\\
&& \ \ \ \  \ \ \ \ \left. + b_3 \left(\phi^*_{1/2}\phi^*_{7/2}+
\phi^*_{3/2}\phi^*_{5/2}\right) \right]\label{psi4_n2}~,
\eea
${\cal N}_0$, ${\cal N}_2$, ${\cal N}_4$ being normalization factors
and $a_1,a_2,a_3,b_2$ and $b_3$ free parameters. 

Of significance is the correspondence between 
the number of parameters in the wave function
(in eq. (\ref{psi0_n2}) the parameter has been absorbed in the 
normalization factor) and the roots degeneracy, namely the number of
independent eigenfunctions of given energy, seniority and zero third 
component of the angular momentum which can be set up with the $\phi$'s
as building blocks.

A few comments are now in order. 
First, clearly, three distinct eigenvalues are just expected for two pairs, 
corresponding to zero, one and two pairs broken respectively. 
By extension, no matter what the value of $\Omega$ is, the distinct roots 
of the secular equation will always be $n+1$. 
Their degeneracy can be found, as above stated, by inspecting the 
associated wave functions. However it can be understood 
on the basis of the following counting rule: 
the lowest eigenvalue is always unique 
$${\cal E}_0(n) \longrightarrow \delta_0=1$$
and corresponds to the most collective 
eigenvector $\Psi_0$ (see Eqs.\ref{psi0_n1}-\ref{psi0_n2}). 
The second eigenvector $\Psi_2$ has a broken pair: since the states 
available for a pair are 
$\Omega$, the broken pair can sit in any of them, except for the one 
corresponding to $\Psi_0$: hence its degeneracy is 
$${\cal E}_2(n) \longrightarrow \delta_2= {{\Omega}\choose{1}}-1 = \Omega -1~.$$
The next eigenvector $\Psi_4$ has two pairs broken. 
The associated degeneracy is obtained by subtracting, out of 
the states available for two pairs, the states already occupied by 
$\Psi_0$ and $\Psi_2$, namely
$${\cal E}_4(n) \longrightarrow 
\delta_4={{\Omega}\choose{2}}-1 -\left[{{\Omega}\choose{1}}-1\right]=
{{\Omega}\choose{2}} -{{\Omega}\choose{1}}~. $$

In general the following degeneracy for the eigenvalues 
corresponding to seniority $v$ is found to hold
(of course, a binomial coefficient with a negative lower index vanishes)
$$\delta_v=
{{\Omega}\choose{\frac{v}{2}}} -1 -\left[{{\Omega}\choose{1}}-1\right] 
-\left[ {{\Omega}\choose{2}} -{{\Omega}\choose{1}}\right] 
\cdots -\left[{{\Omega}\choose{\frac{v}{2}-1}}- 
{{\Omega}\choose{\frac{v}{2}-2}}\right]$$
\be
{\cal E}_v(n) \longrightarrow \delta_v=
{{\Omega}\choose{\frac{v}{2}}} - {{\Omega}\choose{\frac{v}{2}-1}} \ ,
\label{degenerazione}
\ee
$v$ being an {\it even non-negative number} referred to as {\it seniority}, 
to comply with the existing literature. 

Finally we explore the general case. 
Here the structure of the symmetric matrix of dimension 
${\Omega}\choose{n}$ associated with $H_P^{(j)}$ is easily found to be
\be
\left(\begin{array}{ccc}
{\cal E} +n &  & 0\lor 1\\
&\ddots  & \\
0\lor 1 & & {\cal E} +n  
\end{array}
\right) \ ,
\label{matrice_generica}
\ee
where the symbol $0\lor 1$ means that the upper (and the lower)
triangle of the matrix is filled with zeros and ones.  
Indeed the matrix elements of $H_P^{(j)}$ turn out to be one, when the 
bra and the ket differ by the quantum state of one 
(out of $n$) pair, otherwise they vanish. 
The diagonal matrix elements simply count the number of pairs. 

An elementary combinatorial analysis shows that the number of ones in each row 
(column) of the matrix is given by $n(\Omega -n)$.
Indeed a non vanishing matrix element has the row specified by $n$ 
indices whereas, of the indices identifying the column, $n-1$ should be 
extracted from the $n$ ones fixing the row in all the possible ways, which
amounts to $n$ possibilities. The missing index should then be selected 
among the remaining $\Omega-n$ ones: hence the formula $n(\Omega - n)$ 
for the number of ones. 
 
No matter where these are located {\footnote {As already noticed, any 
ordering of the states should of course lead to the same results. There are 
${{\Omega}\choose{n}}!$ ordering alltogheter. Note however that for $\Omega$
and $n$ large the number of matrix with $n(\Omega -n)$ ones in each row and
column is larger than ${{\Omega}\choose{n}}!$.}}, in the determinant of the 
matrix the first row (or column) can be replaced with the sum of all 
the rows (or columns). 
This leads to a new determinant with, e.g., the first row (or column) 
filled with ones and having the expression ${\cal E} +n +n(\Omega-n)$ 
factorized. Hence 
\begin{enumerate}
\item
the lowest eigenvalue is given by the {\it integer}
\be
{\cal E}= -n(1+\Omega-n) = -n-n(\Omega -n)\ ;
\label{E_lowest}
\ee
\item
it is fixed uniquely by $n$ and by the number of ones 
present in each row (or column) of the matrix. 
\end{enumerate}
Next notice that, setting ${\cal E} =0$ in (\ref{matrice_generica}), 
{\it for a given $\Omega$}, the number of pairs
$n (\le \frac{\Omega}{2})$, beyond fixing
the dimension of the matrix and the number of ones in each row (column),
also represents the common value of the diagonal elements. 
We have numerically checked that
the determinant of such a matrix always vanishes. Hence it follows 
that ${\cal E}=0$ is always an eigenvalue of $H_P^{(j)}$ 
and actually the upper one. 
Indeed it corresponds to the situation where all the pairs are broken. 

Let us now insert into the matrix (\ref{matrice_generica}) 
the lowest eigenvalue (\ref{E_lowest}). 
We obtain
\be
det \left(\begin{array}{ccc}
-n(\Omega-n) &  & 0\lor 1\\
&\ddots  & \\
0\lor 1 & & -n(\Omega-n)  
\end{array}
\right)=0~.
\ee
We extensively checked by direct computation 
that the above determinant continues to vanish under the replacement 
$$ -n(\Omega - n) \longrightarrow -n(\Omega - n) + \frac{v}{2} 
\left(\Omega - \frac{v}{2} +1\right)~.$$
Hence the characteristic equation in the general case turns out to be

\be
\prod_{v/2=0}^n 
\left[{\cal E} + n (\Omega-n+1) -\frac{v}{2} 
(\Omega-\frac{v}{2}+1) \right]^{\delta_v} = 0
\ee 
and the whole spectrum of the pairing hamiltonian follows. 

The associated wave functions will be linear combinations of 
${\Omega}\choose{n}$ monomials, each monomial being the product of $n$ 
$\phi$'s, as usual.

In conclusion, we like to shortly address the problem of the angular 
momentum of our eigenstates, which belong to a definite value of 
energy, seniority and third component of the angular momentum, 
but not of the angular momentum. 
However, their building blocks, the monomials, can be expressed as
superpositions of ``collective'' bosons $\Phi_J$ of definite angular 
momentum, according to
\be
\phi_m = \sqrt 2 \sum_J \langle jm,j-m | J0\rangle \Phi_J~.
\label{phiPhi}
\ee
It should be realized however that the nihilpotency of the variables $\phi$'s
induces a set of constraints to be fulfilled by the ``collective'' variables
$\Phi_J$. As an example, consider again the case $\Omega=4,~n=2$. 
Here the four constraints $\phi_m^2=0$ are translated in the following 
four equations for the variables $\Phi_J$ with $J=0,2,4$ and $6$:
\bea
&&\!\!\!\!\!\!\!\!\!\!\!\!\!\Phi_0^2= -\Phi_2^2 -\Phi_4^2 -\Phi_6^2
\label{A2}\\
&& \!\!\!\!\!\!\!\!\!\!\!\!\!
77\sqrt{21} \Phi_0\Phi_2= -88 \Phi_2^2 -10 \Phi_4^2 + 98 \Phi_6^2 -
   44 \sqrt{33} \Phi_2 \Phi_4 - 35 \sqrt{21} \Phi_4 \Phi_6
\label{AD}\\
&& \!\!\!\!\!\!\!\!\!\!\!\!\! 
231\sqrt{77} \Phi_0\Phi_4= -726 \Phi_2^2 +432 \Phi_4^2 + 294 \Phi_6^2 -\nn\\
&&~~~~~~~~~~~   20 \sqrt{33} \Phi_2 \Phi_4 - 105 \sqrt{77} \Phi_2 \Phi_6
+ 280 \sqrt{21} \Phi_4 \Phi_6
\label{AG}\\
&& \!\!\!\!\!\!\!\!\!\!\!\!\! 
11\sqrt{33} \Phi_0\Phi_6= 20 \Phi_4^2 -20 \Phi_6^2 -
   5 \sqrt{33} \Phi_2 \Phi_4 - 4 \sqrt{77} \Phi_2 \Phi_6 
  + 4 \sqrt{21} \Phi_4 \Phi_6\,.
\label{AI}
\eea

Hence the $v=0$ eigenstate (\ref{psi0_n2}), when expressed through the
$\Phi_J$, reads
\be
\Psi_0(n=2) = {\cal N}_0 \frac{1}{2}\left(-3 \Phi_0^2 + \Phi_2^2 
+ \Phi_4^2 + \Phi_6^2\right) 
\label{psi0_Phi}
\ee 
which, exploiting (\ref{A2}), actually becomes 
\be
\Psi_0(n=2) = -2{\cal N}_0 \Phi_0^2~.
\label{PHI0^2}
\ee
The above is the product of two $s$-bosons, as the variable $\Phi_0$ is often 
referred to, or of the wavefunctions of two unbroken pairs. 

Analogously the $v=2$ states, exploiting the constraints (\ref{AD}-\ref{AI}),
turns out to read
\bea
&&\Psi_2(n=2) = 
{\cal N}_2 \frac{2}{\sqrt 3 \sqrt 7 \sqrt {11}} \left[\left(
4a_1-2a_2-a_3\right)\sqrt {11} \Phi_0\Phi_2 + \right.\nn\\
&& + \left.\left( -3a_1 -2a_2 -8a_3\right)\sqrt 3 \Phi_0 \Phi_4 + 
\left(-2a_1-5a_2+2a_3\right)\sqrt 7 \Phi_0\Phi_6\right]\,,
\eea 
which shows that a particular choice of the parameters allows to 
select a state of good angular momentum $(J=2,4,6)$.
Hence this state can be reduced to the product, e.g., of an $s$ and $d$ boson,
the latter corresponding to the wave function of the broken pair. 

More subtle is the $v=4$ case. Here the procedure above outlined leads to the
following expression of the wavefunction
\bea
&&\Psi_4(n=2) = {\cal N}_4 \cdot\nn\\
&&\left[b_2\left( -\frac{4}{7} \Phi_2^2 - \frac{5}{77} \Phi_4^2 +
\frac{7}{11} \Phi_6^2 + \frac{32}{7 \sqrt{33}} \Phi_2 \Phi_4 + 
\frac{36}{3 \sqrt{77}} \Phi_2 \Phi_6 + \frac{8}{11 \sqrt{21}} \Phi_4 \Phi_6
\right)\right. \nn\\
&&\left.+ b_3 \left(\frac{5}{7} \Phi_2^2 - \frac{5}{7} \Phi_4^2 -
 \frac{40}{7 \sqrt{33}} \Phi_2 \Phi_4 - 
\frac{4}{\sqrt{77}} \Phi_2 \Phi_6 + \frac{4}{\sqrt{21}} \Phi_4 \Phi_6
\right)\right]\label{v4Phi}
\eea  
where, remarkably, the collective $\Phi_0$ does not appear, as it should, 
since (\ref{v4Phi}) is associated with two broken pairs.

\end{document}